\documentclass[fleqn,usenatbib]{mnras}

\usepackage{newtxtext,newtxmath}

\usepackage[T1]{fontenc}

\DeclareRobustCommand{\VAN}[3]{#2}
\let\VANthebibliography\thebibliography
\def\thebibliography{\DeclareRobustCommand{\VAN}[3]{##3}\VANthebibliography}

\usepackage{graphicx}	
\usepackage{amsmath}	

\title[Low-frequency radio transients in LoTSS and TGSS]{Limits on long-timescale radio transients at 150 MHz using the TGSS ADR1 and LoTSS DR2 catalogues}

\author[I. de Ruiter et al.]{
Iris de Ruiter$^{1}$\thanks{E-mail: i.deruiter@uva.nl},
Guillaume Leseigneur$^{2}$,
Antonia Rowlinson$^{1,3}$,
Ralph A.M.J. Wijers$^{1}$,
\newauthor
Alexander Drabent$^{4}$,
Huib T. Intema$^{5}$
and Timothy W. Shimwell$^{3,5}$
\\
$^{1}$Anton Pannekoek Institute for Astronomy, University of Amsterdam, Science Park 904, 1098 XH Amsterdam, The Netherlands\\
$^{2}$MAUCA — Master track in Astrophysics, Université Côte d'Azur \& Observatoire de la Côte d'Azur, Parc Valrose, 06100 Nice, France\\
$^{3}$ASTRON, the Netherlands Institute for Radio Astronomy, Postbus 2, NL-7990 AA Dwingeloo, the Netherlands\\
$^{4}$Thüringer Landessternwarte (TLS), Sternwarte 5, 07778 Tautenburg, Germany\\
$^{5}$Leiden Observatory, Leiden University, P.O. Box 9513, NL-2300 RA Leiden, the Netherlands\\
}

\date{Accepted XXX. Received YYY; in original form ZZZ}

\pubyear{2021}

\begin{document}
\label{firstpage}
\pagerange{\pageref{firstpage}--\pageref{lastpage}}
\maketitle

\begin{abstract}
We present a search for transient radio sources on timescales of 2-9 years at 150 MHz. This search is conducted by comparing the first Alternative Data Release of the TIFR GMRT Sky Survey (TGSS ADR1) and the second data release of the LOFAR Two-metre Sky Survey (LoTSS DR2). The overlapping survey area covers 5570 $\rm{deg}^2$ on the sky, or 14\% of the total sky. We introduce a method to compare the source catalogues that involves a pair match of sources, a flux density cutoff to meet the survey completeness limit and a newly developed compactness criterion. This method is used to identify both transient candidates in the TGSS source catalogue that have no counterpart in the LoTSS catalogue and transient candidates in LoTSS without a counterpart in TGSS. We find that imaging artefacts and uncertainties and variations in the flux density scales complicate the transient search. Our method to search for transients by comparing two different surveys, while taking into account imaging artefacts around bright sources and misaligned flux scales between surveys, is universally applicable to future radio transient searches. No transient sources were identified, but we are able to place an upper limit on the transient surface density of $<5.4 \cdot 10^{-4}\ \text{deg}^{-2}$ at 150 MHz for compact sources with an integrated flux density over 100 mJy. Here we define a transient as a compact source with flux density greater than 100 mJy that appears in the catalogue of one survey without a counterpart in the other survey.

\end{abstract}




\section{Introduction}
There are several astrophysical phenomena that are known to be transient at low frequencies. These include events like stellar flares, magnetar flares, intermittent pulsars and X-ray binaries. See Section 4 of \cite{bowman2013science} for a review of the low-frequency transient radio sky. In this study we focus on searching for previously unknown low-frequency (150 MHz) long-timescale (> year) radio transients and extreme variables. Our search is sensitive to various phenomena, for example, active galactic nuclei (AGNs) are known to be variable on these timescales (Figure 3 in \cite{pietka2015variability}) and at these frequencies \citep{hajela2019gmrt}. Both variable radio AGN \citep{williams2016no, nyland2020variable, kunert2020caltech} and changing-look AGN \citep{wolowska2017changing, wolowska2021caltech} may be observed as transient radio emission. Intrinsic AGN variability can arise due to variations in accretion rate, flares and shocks in disks and jets, transitions between high and low states, changes in Doppler boosting and jet precession, and other processes taking place near the black hole (see \cite{bignall2015time} and references therein).

More recently, it has been found that tidal disruption events can have long lasting and detectable radio afterglows \citep{vanVelzen2015radioTDE, de2019evidence, tingay2020archival, anderson2020caltech, ravi2021first}. Furthermore, the afterglow of a gamma-ray burst (GRB) \citep{van2008detailed, chandra2012radio} and neutrons star mergers can be seen up to decades after the event \citep{metzger2015extragalactic}. These commonly searched for explosive events will generally have low flux densities at low frequencies \citep{metzger2015extragalactic}, and we do not expect to be sensitive to them in this study. For the same reason we do not expect to probe core-collapse supernovae. Next to intrinsic variability, a radio source may seem transient or variable because of propagation effects such as refractive scintillation (rather than diffractive scintillation) \citep{rickett1986refractive}. Refractive scintillation is caused by large scale gradients in the interstellar density profile, which can alter the observed flux density and location of compact background sources if not taken into account properly \citep{spangler1989role, goodman1997radio, stinebring2000five}. 

Transient searches at 330 MHz have yielded multiple detections, including an X-ray binary, an unclassified variable source varying on long timescales \citep{hyman2002low} and a rapidly varying source with flares on a minute timescale \citep{hyman2005powerful, hyman2009gcrt, spreeuw2009new}.  At 325 MHz, \cite{jaeger2012discovery} discovered multiple variable sources and a day-scale transient event with no apparent astronomical counterpart. They conclude that this event is likely due to coherent emission from a stellar flare. Short-timescale stellar flares (see for example \cite
{lynch2017154}) will not be detected in this study but the long-term variability that is also found in these systems \citep{callingham2021low} might be found in this long-time scale study. At 150 MHz variable sources have been detected by \cite{sabater2021lofar}. Examples of a transient sources at low frequency include even shorter timescale transients, see for example \cite{stewart2016lofar}, \cite{varghese2019detection} and \cite{kuiack2020long}, who find transients at 34 and 60 MHz. 

Apart from AGN and long-term variability in flare stares, the aforementioned studies are examples of low-frequency variables or transients which are most likely to be detected with repeated observations on short timescales. Examples of long-timescale transients are more sparse.  \cite{law2018discovery} find a transient at 1.4 - 3.0 GHz fading over 23 years. They interpret it as a synchrotron blast wave of a long GRB. The aforementioned studies are mainly examples of blind transient searches, but there are more examples of radio surveys at low frequencies targeted at for example microquasar V404 Cyg \citep{broderick2015lofar, chandra2017giant}, GW170817 \citep{broderick2020lofar} and (off-axis) GRB \citep{mooley2021late}. The low-frequency long-timescale transient radio sky remains largely unexplored. Most variable and transient phenomena are expected at shorter timescales \citep{radcliffe2019insight}, but comparing two readily available large radio surveys such as TGSS ADR1 and LoTSS DR2 still probes an interesting range of transient sources in an overlapping survey sky area of 5570 $\rm{deg}^2$. The most important example to highlight the discovery potential comes from the work by \cite{murphy2017search}, who find a 304 mJy transient of unknown origin on a 3 year timescale at 147.5 MHz, which roughly matches the timescale and frequency of this work. 

Next to the possible identification of transient phenomena this study will put more stringent constraints on the transient rates. Two decades of research have consistently improved the constraints on the transient surface density at various flux density limits, timescales and frequencies. See for example Figure 6 in \cite{anderson2019new} or Figure 8 in \cite{murphy2017search} for an overview of the transient surface density space. The latter figure makes a clear distinction between the timescales of the transient searches. As can be seen from this figure transient searches with timescales of over a year have only been conducted by \cite{rowlinson2016limits} and \cite{murphy2015limits} at timescales of 1 and 3 year respectively. \\
In this paper we present a search for radio transients at 150 MHz, over timescales of 2 to 9 years, to an integrated flux density cutoff of 100 mJy. This search is conducted by comparing the first Alternative Data Release of the TIFR GMRT Sky Survey (TGSS ADR1; \cite{intema2017TGSSADR}) at 150 MHz and the second data release of the LOFAR Two-metre Sky Survey (LoTSS DR2; \citeauthor{shimwell2021DR2} in prep.) at 144 MHz.

\section{Methods}
\subsection{Description of the surveys}
\begin{figure}
    \centering
    \includegraphics[width=\linewidth]{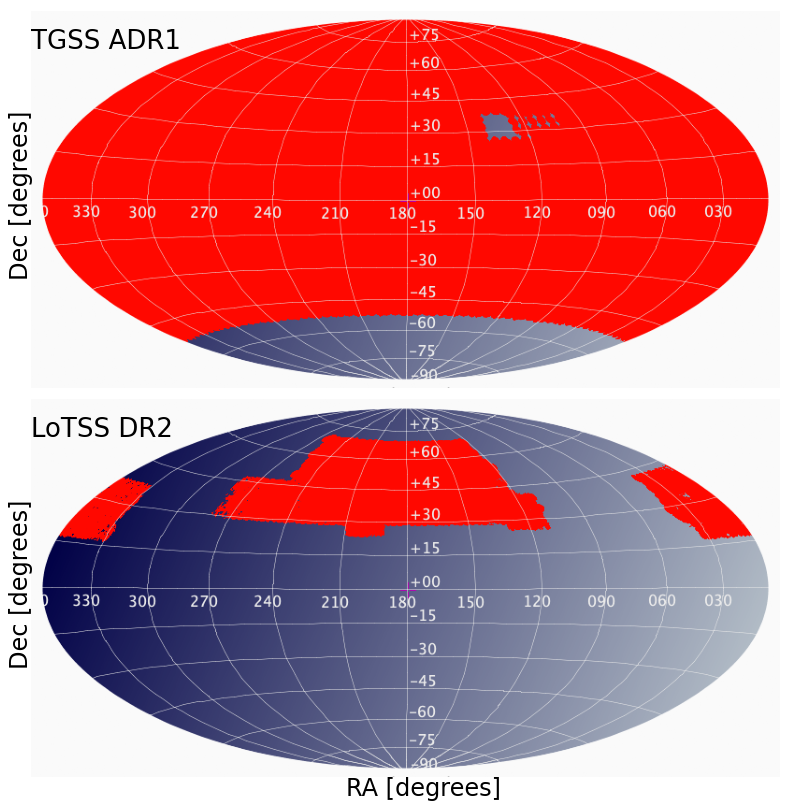}
    \caption{Sky coverage of TGSS ADR1 (top) and LoTSS DR2 (bottom). The red areas indicate the sky covered by the respective surveys. The TGSS survey covers $36900 \; \rm{deg}^2$ or 90\% of the total sky \protect \citep{intema2017TGSSADR}. The LoTSS survey consist of two fields, labeled the 0h and 13h field based on their right ascension. The LoTSS survey covers $5720 \; \rm{deg}^2$ or 14\% of the total sky \protect (\citeauthor{shimwell2021DR2} in prep.) . 
    The missing coverage at low declination in TGSS is due to the declination limit of GMRT. The LoTSS survey is ongoing and will eventually cover the entire northern sky. The missing pointings (holes in the sky coverage) in both surveys are due to observing sessions with very difficult ionospheric conditions.  This representation was created using Aladin \protect \citep{bonnarel2000aladin}.}
    \label{fig:sky_coverage}
\end{figure}

The Giant Metrewave Radio Telescope (GMRT; \cite{swarup1991giant}) is an array of 30 45-meter antennas near Pune, India. GMRT operates at frequencies between 150 and 1500 MHz. The maximum baseline is 25 km. The TGSS survey was carried out between April 2011 and March 2012. \cite{intema2017TGSSADR} have reprocessed the TGSS observations and produced the TGSS Alternative Data Release 1 (TGSS ADR1), 
which covers 36900 square degrees on the sky north of $\delta = -53 ^\circ$ or 90\% of the total sky (see Fig. \ref{fig:sky_coverage}). TGSS ADR1 contains the Stokes I continuum images and a source catalogue containing 623604 radio sources, down to the $7\sigma$ level. The catalogue reaches 90\% point-source completeness at 100 mJy. 
The continuum images have a median RMS of 3.5 mJy $\rm{beam}^{-1}$ and a resolution of $25 \arcsec$. Due to the uv-coverage of the observations, the resolution drops off below declinations south of $\delta = +19 ^\circ$. These declinations are not considered in this work, as LoTSS DR2 only contains regions above this declination.

The Low Frequency Array (LOFAR; \cite{vanhaarlem2013lofar}) is comprised of many thousands of dipole antennas arranged in stations. These stations are distributed in a sparse array with a denser core region near Exloo, the Netherlands, but extending out to remote international stations. The Dutch stations give a maximum baseline of 121 km, while the full international array gives a maximum baseline of 2000 km \citep{vanhaarlem2013lofar}. 
Whilst LOFAR Two-Metre Sky Survey (LoTSS; \cite{shimwell2017lofar}) observations are taken using the full international array, the LoTSS DR2 utilises data only from the Dutch stations. LoTSS observes between 120 and 168 MHz. The flux densities are given at the central frequency of 144 MHz. The survey will eventually cover the entire northern sky. The sky coverage of each data release increases for the ongoing survey \citep{shimwell2017lofar, shimwell2019LoTSS}. The second data release of LoTSS (LoTSS DR2; \citeauthor{shimwell2021DR2} in prep.) consists of two discrete fields, denoted the 0h and 13h fields, covering 5720 square degrees in total (see Fig. \ref{fig:sky_coverage}). Observations for this data release have been made between May 2014 and February 2020. The continuum images have a median RMS of 83 $\mu Jy \; \rm{beam}^{-1}$ and a resolution of $6 \arcsec$. The LoTSS DR2 source catalogue contains 4395448 sources and is 90\% point-source complete at an integrated flux density of 0.8 mJy.
The sensitivity difference between the two surveys is reflected in the number of sources in both catalogues. In the overlapping survey area LoTSS DR2 contains $4.3\cdot 10^6$ sources while TGSS ADR1 contains $1.0 \cdot 10^5$ sources.

\subsection{Search strategy} \label{sec:methods_strategy}
`Transients' are often thought of as sources with a large amplitude
of variability, that undergo brightenings only rarely, or even only once  (cataclysmically). This is then taken in contrast with `variables', thought of as objects that are usually present, but vary in brightness with relatively smaller amplitudes. However, this is a distinction we cannot generally make with discovery data such as in this study: by the nature of flux density distributions, most sources in a survey are close to the detection limit, including new sources. We can only learn by followup observations whether a new source we find is usually just below the detection threshold, or far below, and thus whether its outbursts have high or low amplitude and whether they are rare or common (e.g., \cite{murphy2017search}). 
In this study, we therefore adopt the purely pragmatic definition that a
transient is a source that definitely appears in one catalogue, and is without a
counterpart in the other. Radio surveys, by their nature, are constructed
with a certain flux or completeness limit, which is set such as to have a reasonable
balance between false positives (included sources that are artefacts or noise) and false negatives (sources above the nominal threshold that are missed); inevitably
neither is zero, since the goals of surveys are mostly statistical. However,
transients are quite rare, and thus when trying to find them by comparing two
radio catalogues, the false positives and false negatives in them dominate the
initial difference list. Therefore, much of the work in this study concerns
finding stricter cuts on the catalogue data, by source type and source flux, that ensure
sufficiently smaller error rates that we can call any differences in the
restricted sample new or transient with reasonable confidence. 

We conduct our search by comparing the TGSS ADR1 \citep{intema2017TGSSADR} and LoTSS DR2 (\citeauthor{shimwell2021DR2} in prep.) surveys. The sky area covered by LoTSS DR2 is a subset of the TGSS ADR1 survey, as shown in Fig. \ref{fig:sky_coverage}. Except for a small patch around $(\alpha, \delta) \sim (135, +30) ^\circ$  the total search area is simply defined by the total LoTSS DR2 area. Using these two surveys, one can make two comparisons to identify transients: compare all TGSS sources to the LoTSS sources and vice versa. 
This section describes a  search strategy that applies to both the LoTSS to TGSS as the TGSS to LoTSS comparison. The steps are as follows:
\begin{itemize}
    \item \textbf{Pair match} - We use the TOPCAT \citep{taylor2005topcat} sky ellipses pair match algorithm  with a search radius of $1.0 \arcmin$ to cross-match the catalogues. This algorithm compares the elliptical regions on the sky around all sources for overlap. Sources whose sky ellipse overlap are a match. In case there are multiple sources within the sky ellipse of one source, we select the "best" match. The \textit{Best match for each Table 1 row} output option implies that for each entry in one catalogue, only the best match from the other catalogue will appear in the result. Here "best" match means "closest" - the match with the smallest distance between the two matched celestial positions along a great circle.\footnote{from  \url{http://www.star.bris.ac.uk/~mbt/topcat/sun253/matchRowSelect.html}} If we search for matches for all TGSS sources, each TGSS source will appear a maximum of once in the result, but sources from LoTSS may appear multiple times (or the other way around for the LoTSS to TGSS search). The join type determines which entries are shown in the resulting table based on a match in the catalogues. Here, we want a result table showing all sources in the first catalogue that are not matched to a source in the second catalogue. Therefore, the join type was set to \textit{1 not 2}, resulting in a catalogue of TGSS sources without a match in LoTSS, or vice versa. 
    \item \textbf{Flux density cutoff} - We apply a flux density cutoff of 100 mJy on the sources without a match. This corresponds to the TGSS completeness limit \citep{intema2017TGSSADR} for the entire TGSS ADR1 survey. By setting this flux density cutoff we prevent ourselves from interpreting sources below the completeness limit as false positive transient candidates. We experimented with a flux density cutoff of 50 mJy but find that the number of transient candidates left for visual inspection is almost 20 times higher for the LoTSS to TGSS search and about 2 times higher for the TGSS to LoTSS comparison. We settle on a flux density limit of 100 mJy to prevent enormous visual inspection tasks (with their own risk of error) for a relatively small scientific gain, as a 50 mJy limit is not significantly more constraining than a 100 mJy limit in the transient surface density phase space (see the right hand side of Figure \ref{fig:transient_rates}). 
    \item \textbf{Compact sources} - We select only compact sources from the remaining sample. Based on light travel-time arguments, extragalactic sources that vary significantly on timescales of years are necessarily compact. This step is further explained in Section \ref{sec:methods_compactness}.
    \item \textbf{Visual inspection} - We perform a visual inspection of the remaining sources. Here we get rid of sources that were misidentified by the source finding algorithm, or sources that were right on the border of the coverage of one of the surveys.
\end{itemize}

Table \ref{tab:nof_sources_per_step} shows the number of transient candidates left after each of the aforementioned steps. The first row shows the number of transient candidates in the LoTSS to TGSS comparison. The second row shows the number of transient candidates in the TGSS to LoTSS comparison.

\begin{table}
\begin{tabular}{cccccc}
      & Total            & No match & > 100 mJy   & Compact          & Visual \\ \hline
LoTSS & $4.3 \cdot 10^6$ & $2.6 \cdot 10^6$       & $2.0 \cdot 10^3$ & 60               & 16                \\
TGSS & $1.0 \cdot 10^5$ & $4.0 \cdot 10^2$       & $1.5 \cdot 10^2$ & $1.2 \cdot 10^2$ & 10               
\end{tabular}
\caption{\label{tab:nof_sources_per_step} Number of sources that is left after each search step as explained in Section \ref{sec:methods_strategy}. The total number of sources refers to the number of sources in each survey in the overlap region of the two surveys.}
\end{table}

\subsection{Compactness \label{sec:methods_compactness}}
To select compact sources we follow a similar strategy as in Section 3.1 and Figure 7 in \cite{shimwell2019LoTSS}. We also tried a strategy as in Figure 11 \cite{intema2017TGSSADR} to select compact sources but find that this is too strict for transient searches. We would rather have a few more sources to visually inspect than to disregard possible transient candidates too soon. In the future, as survey resolution and sensitivity increase, a more stringent compactness constraint might be necessary. This section is similar to methods in \citeauthor{shimwell2021DR2} (in prep.) but we try to create a standard procedure that can be applied to both LoTSS DR2 and TGSS.

In short, we define compact sources by fitting an envelope to a preselected sample of compact sources. The criteria for these preselected compact sources are listed below. Finally, a fit is performed on the preselected compact sources that defines the boundary between compact and extended sources. Any source of the full catalogue (not just the subsample of preselected compact sources) that lies below this envelope is defined as compact. 
The criteria to find a sample of compact sources to perform the fit on are as follows. From the full catalogue:
\begin{enumerate}
    \item Select sources that are classified as "S" in the catalogue by the source finder (PyBDSF \citep{mohan2015pybdsm} for both catalogues). This source code corresponds to sources that are well fit by a single Gaussian.
    \item Select sources with an integrated flux density greater than the completeness limit of the survey. 
    \item Select sources without other catalogue sources within $1\arcmin$.
    \item Select sources with a major axis smaller than two times the restored beam size. The restored beam size is $6\arcsec$ for LoTSS and $25\arcsec$ for TGSS (above $\delta = 20 ^\circ$).
\end{enumerate}
Step (ii) differs from the strategy in \cite{shimwell2019LoTSS} where they instead exclude sources that were not in the deconvolution mask in every pointing in which they are detected. We instead replace this requirement with a threshold on the integrated flux density, which in the end results in a similar compactness envelope. A constraint on the integrated flux density can also easily be applied to TGSS. Step (iv), the constraint on major beam size, is based on \cite{shimwell2019LoTSS} where a limit of $15\arcsec$, i.e. 2.5 times the restored beam size, was chosen. We change this to 2 times the restored beam size as this seems to work well for both TGSS and LoTSS. A more conservative approach is to choose 2.5 times the restored beam size, which will result in a larger number of sources to visually inspect. We try both 2.5 and 2 times the restored beam size and find that it does not make a difference in the final number of transient candidates. It is important to have a constant restoring beam size throughout the survey for this requirement to make sense. For TGSS sources we thus add an additional requirement on the declination, $\delta > 20 ^\circ$, as the restored beam size increases rapidly below this declination, which would interfere with the final condition. Note that we include all TGSS ADR1 sources to construct a compactness envelope, not just the sources that lie in the LoTSS DR2 field. 

\begin{figure}
    \centering
    \includegraphics[width=\linewidth]{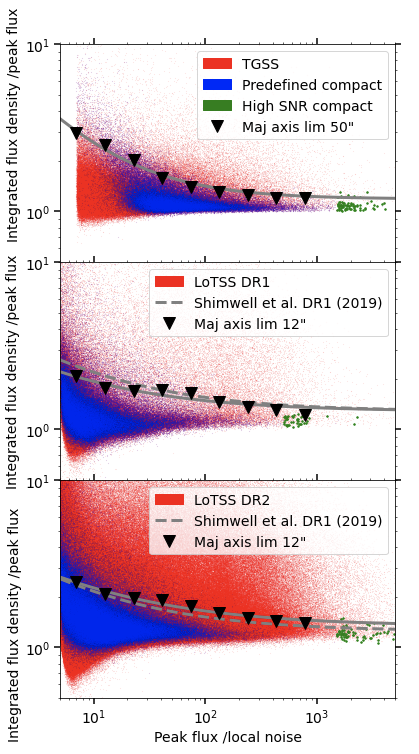}
    \caption{
    Ratio of the integrated flux density to peak brightness as a function of S/N for sources in the TGSS, LoTSS DR1 and LoTSS DR2 catalogue. All catalogued sources are shown in red and the sources we used to define a compact envelope in blue. The green points are high S/N sources used to determine the offset in the fit (Eqn. \ref{eqn:compactness_fit}), which is shown in grey. The downwards pointing triangles encompass 95\% of predefined compact sources in each horizontal bin. A more detailed description is given in Section \ref{sec:methods_compactness}. The fit parameters (Eqn. \ref{eqn:compactness_fit}) of the compactness envelopes in grey:\\
    \begin{tabular}{llll}
                               & Offset & A    & B     \\
    TGSS (top)                 & 1.19   & 8.39 & -0.78 \\
    Shimwell et al. (2019) DR1 (dashed) & 1.25   & 3.1  & -0.53 \\
    LoTSS DR1 (middle)         & 1.28   & 2.10 & -0.51 \\
    LoTSS DR2 (bottom)         & 1.34   & 2.79 & -0.48
    \end{tabular}
    }
    \label{fig:compactness_combined}
\end{figure}

We plot the ratio of the integrated flux density to peak flux as a function of the ratio of the peak flux to the local noise for each source, as in Figure \ref{fig:compactness_combined}.
Figure \ref{fig:compactness_combined} shows the TGSS, LoTSS DR1 and LoTSS DR2 respectively. Since the criteria to find a sample of compact sources differ from the strategy in \cite{shimwell2019LoTSS}, we reanalyse the same source catalogue (LoTSS DR1) to verify our compactness criterion for the TGSS and LoTSS DR2 surveys. The red points show the full source catalogues. The blue points show the predefined compact sample, constructed following the steps above. To separate extended from compact sources we fit an envelope with the functional form
\begin{equation}
    \frac{S_{\rm{int}}}{S_{\rm{peak}}} = \rm{offset} + \mathit{A} \cdot \left( \frac{\mathit{S}_{\rm{peak}}}{\rm{rms}} \right)^{\mathit{B}}
    \label{eqn:compactness_fit}
\end{equation}
that encompasses 95\% of the predefined compact sources. To this end, we define 10 peak flux to local noise (horizontal) bins and for each bin find the integrated to peak flux value that holds 95\% of the predefined compact sources. These 95\% values are shown with the downwards pointing triangles and are fit by the grey line. The offset in this envelope fit is defined by the median plus three times the median absolute deviation of the integrated flux density to peak flux ratio of seemingly compact high S/N sources. For the TGSS survey we impose a signal to noise ratio of 1500. For the LoTSS DR1 and DR2 surveys we impose a signal to noise ratio of 500. We visually inspect these high S/N sources to make sure that they are truly compact and that there are no obvious artefacts. The high S/N compact sources are shown in green in Fig. \ref{fig:compactness_combined}. The envelope fit parameters are given in the plot caption. An additional dashed grey line is shown that represents the compactness envelope from \cite{shimwell2019LoTSS}. Comparing the two envelopes shows that especially for high peak flux to local noise values they are almost identical. This is confirmed by the fact that the compactness envelope established in this work encompasses 83\% of all LoTSS DR1 sources, while the \cite{shimwell2019LoTSS} envelope encompasses 86\% of all LoTSS DR1 sources. We therefore conclude that the criteria to find compact sources, as defined in this section, are suitable to filter out compact sources and we apply these criteria to both the TGSS and LoTSS DR2 survey.
The compactness envelope for LoTSS DR2 encompasses 81\% of all LoTSS DR2 sources. The compactness envelope for TGSS encompasses 91\% of all TGSS sources. Although the percentages given here might not seem very constraining, Table \ref{tab:nof_sources_per_step} shows that the compactness criterion is crucial to reduce the number of transient candidates to an amount that can be visually inspected.

\subsection{Flux density scale} \label{sec:methods_flux_scale}
\begin{figure}
    \centering
    \includegraphics[width=\linewidth]{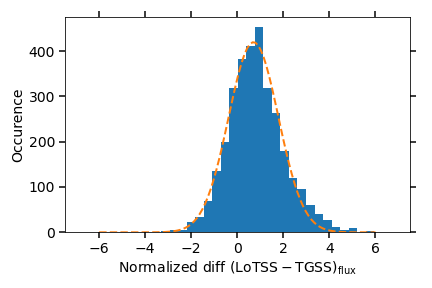}
    \caption{Flux density difference normalized with the root of the squared sum of the errors (as defined in Eqn. \ref{eqn:flux_diff}). The median flux density difference is 0.70 and the standard deviation is 1.08. We overplot the best fitting Gaussian to show the asymmetry in the distribution.
    }
    \label{fig:scaled_flux_diff}
\end{figure}

To make sure that we can compare the flux densities of the TGSS and LoTSS survey, we should check that the flux density scales are properly aligned. Following the previous section we select the compact sources in LoTSS that meet the following condition (See section \ref{sec:methods_compactness} and Figure \ref{fig:compactness_combined})
\begin{equation}
    \frac{S_{\rm{int}}}{S_{\rm{peak}}} < 1.34 + 2.79 \left( \frac{S_{\rm{peak}}}{\rm{rms}} \right)^{-0.48}
\end{equation}
and are matched to a compact source in TGSS. Since extended emission is more likely to be fit by multiple 'sources' in the source finding algorithms, we select compact sources only, to make an accurate flux comparison. After selecting the compact sources, we select sources with a flux density over 100 mJy to match the TGSS completeness limit \citep{intema2017TGSSADR}. We find 6097 compact LoTSS sources that are matched to a compact TGSS source brighter than the completeness limit. This small subset of the total number of sources is sufficient to estimate whether the flux density scales are aligned. Figure \ref{fig:scaled_flux_diff} shows the LoTSS to TGSS normalized flux density difference, defined as 
\begin{multline}
    \rm{Norm. \; diff \; (LoTSS - TGSS)}_{\rm{flux}} =\\ \frac{S_{\rm{int \; LoTSS}} - S_{\rm{int \;  TGSS}}}{\sqrt{(\Delta S_{\rm{int \; LoTSS}})^2 + (\Delta S_{\rm{int \; TGSS}})^2}}
    \label{eqn:flux_diff}
\end{multline}

where $S_{\rm{int \; LoTSS}} $ and $S_{\rm{int \;  TGSS}}$ are the integrated flux densities of the source in the LoTSS and TGSS survey respectively, and $\Delta S_{\rm{int \; LoTSS}}$ and $\Delta S_{\rm{int \; TGSS}}$ are the errors on the integrated flux density. Note that for TGSS $\Delta S_{\rm{int \; TGSS}}$ can be taken directly from the catalogue, as it includes both the statistical error on the total flux density measurement as the flux scale uncertainty, while the LoTSS catalogue only shows the statistical error on the total flux density measurement. Therefore, one should add the flux scale uncertainty of 10\% (\citeauthor{shimwell2021DR2} in prep.) in quadrature to the flux density error in the catalogue to obtain $\Delta S_{\rm{int \; LoTSS}}$. When the flux densities for the matched sources in TGSS and LoTSS would be aligned (within the errors), one would expect a normal distribution with a mean of zero and a standard deviation of one. Figure \ref{fig:scaled_flux_diff} shows that the Gaussian fit to the histogram peaks at 0.70 and the standard deviation is 1.08. Hence, the LoTSS sources have a systematically higher flux density. The ratio between the total integrated flux density of LoTSS and TGSS indicates that LoTSS sources have $\sim 10\%$ higher total flux density. Furthermore, the width of the distribution implies that sources are often further than the sum of the errors apart. The extended tail to the positive end of the distribution in Figure \ref{fig:scaled_flux_diff} should be considered when searching for transient sources that are detected in the LoTSS survey but not in the TGSS survey. A large difference in flux density might cause the transient candidate to drop below the TGSS completeness limit. When searching for transients that are detected in TGSS but not in LoTSS this will be less of a problem because the completeness limit for LoTSS is much lower than for TGSS.
A really small flux density difference is expected due to the frequency difference of the surveys, LoTSS is 144 MHz and TGSS is 150 MHz. However, the wide distribution as shown in Fig. \ref{fig:scaled_flux_diff} implies that the flux scales of  TGSS ADR1 and/or LoTSS DR2 are/is fairly inaccurate. Finally, we note that the normalized flux density difference depends on the sky position position of the source. This effect has to be considered in the analysis of the significance of our transient candidates (Section \ref{sec:results_LoTSS}).

\section{Results}
\subsection{TGSS sources without LoTSS counterpart} \label{sec:results_TGSS}

\begin{figure*}
    \centering
    \includegraphics[width=\textwidth]{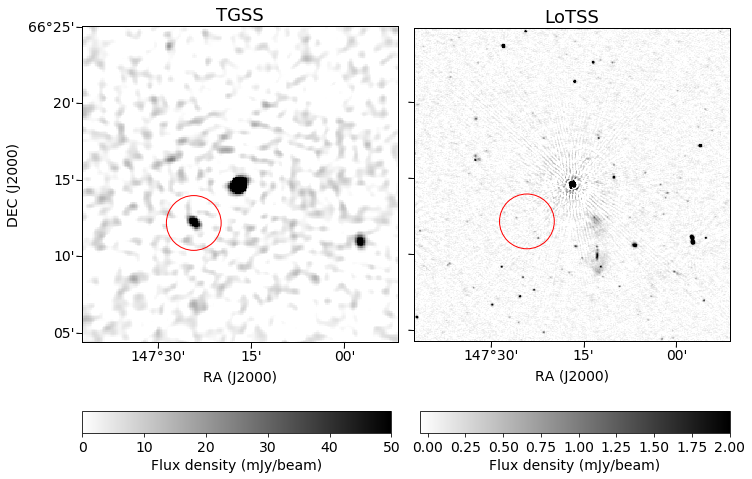}
    \caption{TGSS ADR1 (left) and LoTSS DR2 (right) images of the AGN 4C66.09 shown in the centre of the images. In the red circle shows the location of the possible transient source (TGSSADR J094940.0+661228), visible in the TGSS image.}
    \label{fig:transient_example}
\end{figure*}

The search for transient sources in the TGSS survey leaves us with a sample of 10 transient candidates that are not found in the LoTSS survey. Table \ref{tab:artefact_list} in Appendix \ref{app:TGSS_imaging_artefacts} shows the details of these sources. In Fig. \ref{fig:transient_example} we show an example of a 175.5 mJy transient candidate. The left panel shows the TGSS image, while the right panel shows the LoTSS image. Both panels are centered on AGN 4C66.09, which has a flux density of 4.3 Jy. The location of the transient candidate, visible in TGSS but not in LoTSS, is indicated with a red circle.

We notice that all 10 sources in our final sample lie in close proximity to a bright source. A link to the TGSS and LoTSS images of all brighter than 100 mJy candidates can be found in the reproduction package. Keeping in mind previous studies like \cite{frail2012revised} and \cite{polisensky2016exploring}, that find source-like imaging artefacts around bright sources, we suspect our transient candidates to be imaging artefacts. To further explore the origin of the candidate transient sources in our sample we increase our sample size by lowering the previously set flux density cutoff of 100 mJy to 70 mJy and dropping the compactness criterion. By visual inspection we now find an additional 31 sources which all lie in close proximity to a bright source, as was the case for our initial sample.

\begin{figure}
    \centering
    \includegraphics[width=\linewidth]{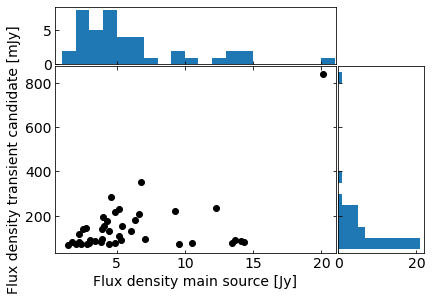}
    \caption{Integrated flux density of the transient candidates as a function of the integrated flux density of the close by main source. This figure includes all transient candidates with an integrated flux density over 70 mJy and without applying a compactness criterion (see Table \ref{tab:artefact_list} in Appendix \ref{app:TGSS_imaging_artefacts}). This figure includes the transient candidates found in the TGSS survey without a LoTSS counterpart.}
    \label{fig:trans_flux_main_flux}
\end{figure}

\begin{figure}
    \centering
    \includegraphics[width=\linewidth]{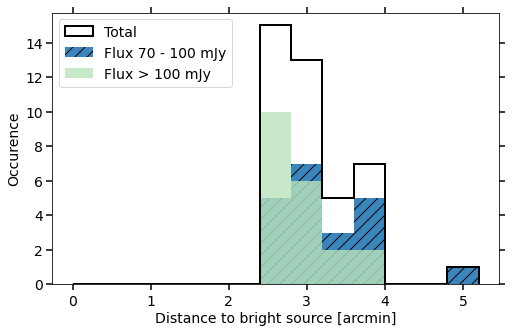}
    \caption{Distribution of distances between the transient candidates and the close by main bright source. We make a distinction between the transient candidates with a flux density over 100 mJy (above the TGSS completeness limit) in green and sources with a flux density between 70-100 mJy that were only found after extending the sample, in striped blue. This figure includes the transient candidates found in the TGSS survey without a LoTSS counterpart.} 
    \label{fig:distance_hist}
\end{figure}

Figure \ref{fig:trans_flux_main_flux} shows the integrated flux density of all 41 sources in the new sample as a function of the close by bright/main source of 1.4 to 20 Jy. The flux density of the transient candidates is between 70 and 850 mJy. 
From Figure \ref{fig:trans_flux_main_flux} we do not find a clear relation between the flux density of the close by bright source and the flux density of the 'transient' source. We can however estimate a linear upper limit where the integrated flux density of the transient candidate is roughly one twenty-fifth of the integrated flux density of the bright source, as might be expected for sidelobes of the point spread function.
Figure \ref{fig:distance_hist} shows the distribution of distances of the sources in our sample to the nearby bright source. The small range of distances of $\sim 2.5 - 4.2\arcmin$ is striking. There is one source (TGSSADR J160357.5+572931) with a distance of $\sim 5.2\arcmin$ to the bright source. This source might be associated with a different bright source at roughly $1.4 \arcmin$.

Reimaging of the TGSS transient candidates with a newer version of the TGSS imaging pipeline shows that most emission at the location of the transient sources with an integrated flux density above 100 mJy (upper half of Table \ref{tab:nof_sources_per_step}) is due to sidelobe artefacts. In the reimaging process the sidelobe artefacts from bright sources are suppressed more effectively than in the original imaging pipeline. We therefore conclude that our TGSS transient candidates are artificial effects of cleaning of a close by bright source. During subtraction of the point spread function a certain amount of flux is mapped to the location of the sidelobes of the point spread function. If this is not properly accounted for, this results in fake sources as seen before in for example \cite{frail2012revised} and \cite{polisensky2016exploring}.

\subsection{Local noise around TGSS artefacts}
The noise properties of the skyregions where the TGSS transient candidates were found are analysed by comparing to the noise properties in the mosaic. All 5 by 5$^\circ$ mosaics are divided in squares of 20 by 20$\arcmin$. In each of these boxes a histogram of pixel values was created and fitted by a normal distribution. The mean and standard deviations of those distributions for all 20 x 20$\arcmin$ boxes are accumulated to form a new distribution. In those distributions of means and standard deviations there was no clear evidence of under- or overcleaning in the boxes were the transient candidates were found. For some transient candidate boxes the mean or the standard deviation of pixel values was at one end of the distribution for the full mosaic, but no systematic behaviour was found.

\subsection{LoTSS sources without TGSS counterpart} \label{sec:results_LoTSS}
As discussed in Section \ref{sec:methods_strategy} we follow the same simple strategy to search for transients sources that are detected in LoTSS but are not detected in TGSS. After applying a flux density threshold of 100 mJy to all LoTSS sources without a TGSS match, we apply the compactness criterion as described in Fig. \ref{fig:compactness_combined}. After visual inspection we are left with 16 (see Table \ref{tab:nof_sources_per_step}) potential transient candidates. Keeping in mind Figure \ref{fig:scaled_flux_diff}, we should design a strategy to evaluate our transient candidates, accounting for (sometimes) large flux density differences in LoTSS and TGSS. The large tail to the right side of this distribution implies that there is a large number of sources where the LoTSS flux density is significantly higher than the TGSS flux density. To this end we perform a forced flux extraction on the location of the transient candidate in the TGSS survey using PySE \citep{carbone2016new}. This allows us to compare the flux density at a particular location between the surveys, although no source has been found by the source extractor in TGSS originally. Using this extracted flux density measurement we evaluate the flux density at the location of the transient candidate in both surveys with respect to other sources in the local environment. The flux density difference between the two surveys at the location of the transient candidate should be compared to the local environment as the flux density difference is dependent on sky position. To this end we construct a figure similar to Figure \ref{fig:scaled_flux_diff} for all transient candidates, where we now only include the 80 sources closest to the location of the transient candidate. These 80 sources are a subsample of the full dataset as described in Section \ref{sec:methods_flux_scale}, they are compact, brighter than 100 mJy sources that are found in both TGSS and LoTSS. We fit this distribution with a Gaussian.

\begin{figure}
    \centering
    \includegraphics[width=\linewidth]{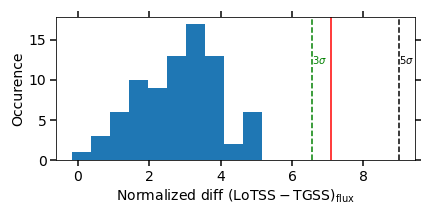}
    \caption{The histogram shows the normalized flux density difference (Eqn. \ref{eqn:flux_diff}) between the two surveys for the 80 compact, brighter than 100 mJy sources around transient candidate ILT J180625.18+385035.1. These sources are used to define the local normalized flux density difference between the surveys. The flux density difference for the transient candidate is shown with a solid vertical red line, which is more than 3 standard deviations away from the local flux density difference distribution.}
    \label{fig:local_noise_hist_example}
\end{figure}

An example of this method is shown in Figure \ref{fig:local_noise_hist_example}. The histogram shows the flux density difference between the two surveys for the 80 compact, brighter than 100 mJy sources around transient candidate  ILT J180625.18+385035.1. These sources are used to define the local normalized flux density difference, $\rm{(LoTSS - TGSS)}_{\rm{flux}}$ as defined in Eqn. \ref{eqn:flux_diff}, between the surveys. For the example in Fig. \ref{fig:local_noise_hist_example} the 80 sources are within $4.7 ^\circ$ of the transient candidate. The flux density difference for the transient candidate is shown with a solid vertical red line, which in this case is more than $3 \sigma$ away from the centre of the Gaussian that was fit to the local distribution.

We apply this method to all 16 transient candidates and find six LoTSS sources where the flux density difference with respect to the TGSS survey is significant to a $3\sigma$ level. The LoTSS sources with a significant flux density difference to the TGSS survey are ILT J164033.71+383905.1, ILT J163248.84+374549.0, ILT J180625.18+385035.1 (Fig. \ref{fig:local_noise_hist_example}), ILT J162817.33+401534.4, ILT J162915.48+655220.9 and ILT J092201.65+312144.8. These sources are reconsidered with a newer version of the TGSS imaging pipeline. It is found that for all sources the flux density increases after reprocessing, which is due to a different choice of primary beam calibrator in the new version of the imaging pipeline. Furthermore, some of the sources lie on the edge of two fields, where a field of poor quality significantly impacts the source properties. This implies that the flux density difference is pushed from above $3\sigma$ (as in Figure \ref{fig:local_noise_hist_example}) to below $3\sigma$ for all sources. We thus conclude that there are no significant transient sources in the LoTSS survey above 100 mJy.

\section{Discussion}

\subsection{Imaging artefacts around bright sources} \label{sec:discussion_artefacts}
\begin{figure*}
    \centering
    \includegraphics[width=\textwidth]{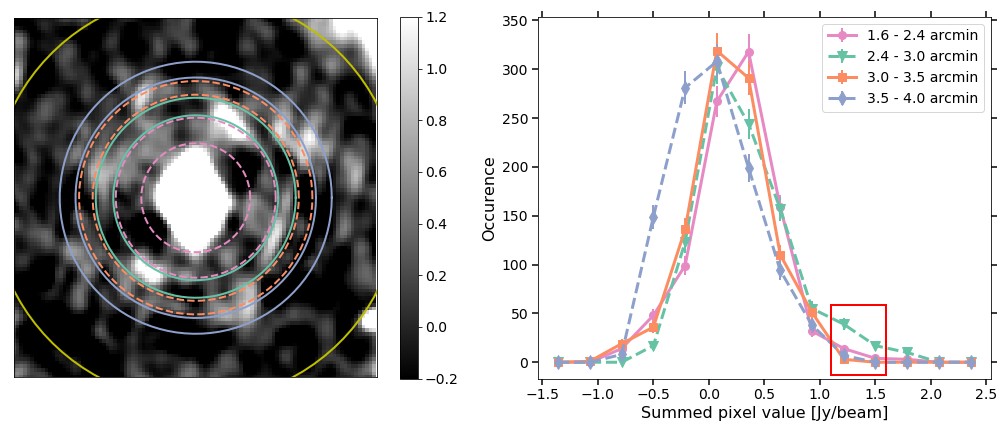}
    \caption{The left panels shows an overlay of all compact (S) TGSS sources with flux density over 2 Jy. We excluded the sources that have another source brighter than 100 mJy within 6 $\arcmin$. Summing  the images of the remaining sources (1144 in total) together to obtain effectively a reconstruction of the residual dirty beam effects. The histograms on the right show the pixel distributions in rings around the image centre. The pixel values of 1.0 to 1.5 Jy/beam occur significantly more often at a distance of $2.4 \arcmin$ to $3.0 \arcmin$ from the image centre, compared to the other distances (see red box).  Furthermore, the pixel values at a distance of $3.5 \arcmin$ to $4.0 \arcmin$ are centered around zero, implying a random distribution and proper cleaning.} 
    \label{fig:dirty_beam_reconstruction}
\end{figure*}

Now that we know that our transient search suffered from imaging artefacts, we try to develop a method to reduce the impact of imaging artefacts in future studies. The easiest way to do so is to create a region around bright sources that is to be excluded from the transient search. This region should be as small as possible as one would like to keep the sky area searched for transients as large as possible. To this end, one can reconstruct residual dirty beam effects that are still in the survey by overlaying all bright sources that are modelled by a single Gaussian (source code S). For the TGSS survey specifically we choose sources with an integrated flux density above 2 Jy, based on Figure \ref{fig:trans_flux_main_flux} and Table \ref{tab:nof_sources_per_step}. After selecting all sources with source code S, we excluded sources that have a catalogued brighter than 100 mJy source within a $6 \arcmin$ proximity. If the cleaning process had been perfect, overlaying all the remaining sources should result in only a bright peak of emission in the centre of the overlayed images. The structure around this bright peak should be random since all sources with a flux density below 100 mJy (that are left) should average out. 

To accumulate better statistics we conduct this analysis with the full TGSS survey (not just the part overlapping with the LoTSS survey). There are 1144 brighter than 2 Jy single Gaussian sources without brighter than 100 mJy sources in $6 \arcmin$ proximity. Figure \ref{fig:dirty_beam_reconstruction} shows the result of overlaying all these sources. The outer solid yellow circle has a $6 \arcmin$ radius, within this circle we do not expect to find significant structure except for a bright central region. However, the image clearly reveals darker and brighter regions close to the image centre. The right side of Figure \ref{fig:dirty_beam_reconstruction} shows a histogram of pixel values in rings at certain distances from the centre of the image. For visual clarity these rings are also indicated on the image on the left side. The distances of 1.6, 2.4, 3.0, 3.5 and $4.0 \arcmin$ are chosen such that there is roughly an equal number of pixels in each ring. The histograms show that at a distance of 2.4 to $3.0 \arcmin$ of the centre (green triangles, dashed line) there is an excess of bright pixels compared to the other distances (see red box). This excess is also visible in the image on the left. The excess of bright pixels at a distance of 2.4 to $3.0 \arcmin$ to the centre is significant compared to the other distances. This distance corresponds perfectly to the peak of the histogram of distances between the TGSS transient candidates and the close by bright sources (Figure \ref{fig:distance_hist}). Furthermore, figure \ref{fig:dirty_beam_reconstruction} shows that starting from 3.5 to $4.0 \arcmin$ the pixel histogram is centered around zero, which implies a random distribution and proper cleaning. Therefore, for the TGSS survey one should exclude regions within $4.0 \arcmin$ of bright sources for the transient search.

\subsection{Transient rates}
Following \cite{rowlinson2016limits} we calculate the transient surface density limit using Poisson statistics via $P = e^{-\rho (N-1)\Omega}$, where $(N-1)\Omega$ is the total area surveyed by $N$ snapshots of a field each with an area of $\Omega$, $\rho$ is the surface density limit and $P$ is the confidence interval. Following \cite{bell2014survey}, we use $P=0.05$ to give a 95 per cent confidence limit. In this work $N=2$ since we are comparing images from 2 surveys, and $\Omega =  5720-150 = 5570 \; \rm{deg}^2$. The sky area is simply the area of the LoTSS DR2 survey minus a small gap in the TGSS ADR1 survey (see Fig. \ref{fig:sky_coverage}). This results in a surface density estimate of $\rho = 5.4 \cdot 10^{-4}\ \text{deg}^{-2} $. This result is only an upper limit since no transients were detected.

\begin{figure*}
    \centering
	\includegraphics[width=\textwidth]{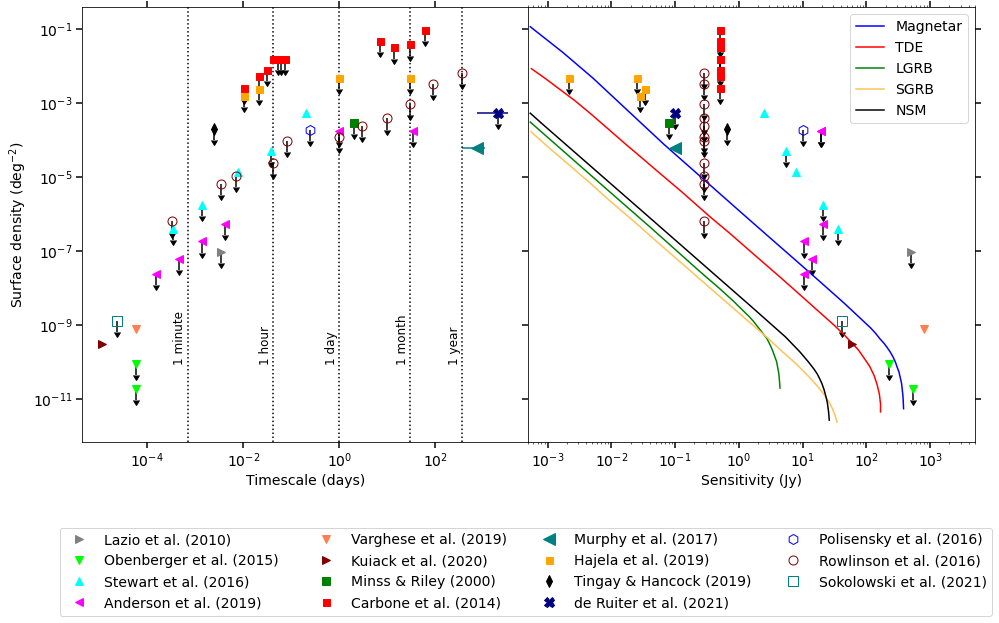}
    \caption{Limits on the transient rates from this study compared to previously published results. The result presented in this paper is shown as a dark blue cross. The horizontal error bar shows the uncertainty in timescale, as the TGSS LoTSS data spans 2 to 9 years. Triangle markers correspond to studies at frequencies below 149 MHz and empty markers correspond to studies at frequencies above 154 MHz, all other markers describe studies conducted between those frequencies. The coloured lines show the sky density of sources above flux density $F_{\nu}$ for frequency $\nu$ = 150 MHz. See \protect\cite{metzger2015extragalactic} for a detailed description of how these model predictions were calculated.
    \protect \citep{lazio2010surveying, obenberger2015monitoring, stewart2016lofar, anderson2019new, varghese2019detection, kuiack2020aartfaac, minns2000compact, carbone2016new, murphy2017search, hajela2019gmrt, tingay2019multi, polisensky2016exploring, rowlinson2016limits, sokolowski2021southern}. Jupyter notebook and supplementary materials are available in the reproduction package. 
    \label{fig:transient_rates}}
\end{figure*}

Figure \ref{fig:transient_rates} shows our new result (navy cross) compared to other results in the literature. We show the most constraining studies below 1 GHz. The structure of this plot was taken from \cite{murphy2017search}, but a more up-to-date sample of the most constraining studies was compiled using an overview\footnote{\url{http://www.tauceti.caltech.edu/kunal/radio-transient-surveys/index.html}} from \cite{mooley2016caltech}. Empty markers correspond to studies at frequencies above 154 MHz and triangle markers correspond to studies at frequencies below 149 MHz, all other markers describe studies conducted in between those frequencies. Markers with a downward pointing arrow represent upper limits. The jupyter notebook used to create this figure is available in the reproduction package. The timescale explored in our study ranges from 2 to 9 years, therefore we plot our surface density at a timescale of 5.5 years with an uncertainty of 3.5 years. Figure \ref{fig:transient_rates} also shows the predicted rates for a range of phenomena, as calculated by \cite{metzger2015extragalactic}. The colored lines show the sky density of sources above flux density $F_{\nu}$ for a frequency of 150 MHz.  We have included predicted source rates for various source classes from Figure 3 in \cite{metzger2015extragalactic} , specifically: magnetars (blue); off-axis tidal disruption events (red); long GRBs with $\theta_{\rm{obs}} = 1.57$ (green); off-axis short GRBs (orange); and neutron star merger leaving black hole (black). For a detailed discussion of these models we refer the reader to \cite{metzger2015extragalactic}, but we would like to point out that the uncertainties associated to these models can be up to an order of magnitude.

Although our study explores the transient sky at longer timescales than ever before, the fact that we only find an upper limit combined with the limited sky area (compared to the \cite{murphy2017search} study) results in a surface density upper limit that is less constraining than previous results at similar sensitivities.

\subsection{Long-term variability in AGN}
Extragalactic radio sources primarily consists of star-forming galaxies and AGN. For the flux densities considered in this work (sources brighter than 100 mJy) the population is dominated by AGN \citep{wilman2008semi, williams2016lofar, calistro2017lofar}. The slow transient radio sky is dominated by AGN emission, the majority of which are likely associated with variability (see for example \cite{nyland2020variable}). 
Assuming that all compact sources brighter than 100 mJy in this study are AGN, we can estimate the number of AGN that show this type of extreme long-term variability. Applying the compactness criterion as described in Section \ref{sec:methods_compactness}, we find $7.1 \cdot 10^3$ compact brighter than 100 mJy sources in LoTSS and $3.9 \cdot 10^4$ compact brighter than 100 mJy sources in TGSS. These numbers show that the compactness criterion is much more restrictive for LoTSS than for TGSS. This leads to a surface density estimate of AGN (at a flux density threshold of 100 mJy and 150 MHz) of $1.3 \ \text{deg}^{-2} $ in LoTSS. Including our previously discussed transient surface density upper limit we estimate a maximum of one in $2.4 \cdot 10^3$ AGN to show extreme long-term variability at 150 MHz above a flux density of 100 mJy. Here we divide the more conservative LoTSS AGN surface density estimate by the previously estimated transient surface density upper limit. Extreme long-term variability in this case means a variable source that can be interpreted as a transient source when comparing LoTSS to TGSS. As mentioned in section \ref{sec:methods_strategy} we can not estimate the amplitude of the variability a source would have to exhibit, for it to be identified as a transient in our search. Since the transient surface density is an upper limit, our estimate of AGN that show long-term variability is a limit as well. A maximum of one in $2.4 \cdot 10^3$ AGN is expected to show extreme long-term variability.

\subsection{Search completeness}
In this section we describe some effects that might have had an impact on the transient search completeness. Some transient sources might have been overlooked because a single source at the resolution of LoTSS could be blended with another source at the resolution of TGSS. This happens when a transient source is close (in sky projection) to a steady source. An extreme example of this is when a transient occurs at the same location as a persistent source (\cite{keane2016host} and follow up discussion by \cite{williams2016no}).

A second possibility is that we have missed transient sources due to false matches between the catalogues. This happens when a transient source has a positional coincidence with a source in the other catalogue. To quantify this effect we conduct a similar study as in \cite{murphy2017search}. We repeat the study as described in Section \ref{sec:methods_strategy} by matching all compact brighter than 100 mJy LoTSS sources to the TGSS sources, after shifting the TGSS sources in a random direction. We add a random offset in right ascension and declination between 5 and $10\arcmin$. This cross-matching yields a match for 37 of the 6651 sources. This implies that up to $\sim 0.6 \%$ of the sources ruled out in our process could have been actual transient phenomena. The total number of transients in our comparison is below one, which means that the expectation is that the number of sources we would miss due to this is less than one. However, many classes of physical transients, such as radio supernovae, will occur at the same location as a persistent radio source (the host galaxy) and this is not accounted for in this analysis.

Finally, we could have missed transient sources if a compact source is shrouded by extended emission. In this case it is not clear if a source finder will correctly identify both the compact source and the extended emission. We investigate this scenario by complementing our initial strategy with a direct comparison of LoTSS sources without a TGSS counterpart to the NRAO VLA Sky Survey \citep[NVSS;][]{condon1998nvss} at 1.4 GHz. The NVSS survey was conducted between 1993 and 1996 and has a median RMS noise of 0.45 mJy $\rm{beam}^{-1}$ and $45 \arcsec$ resolution \citep{condon1998nvss}. We expect that the NVSS survey will detect the compact part of the LoTSS sources without TGSS counterpart, because of the high frequency (where for most radio sources the flux density is higher) and low RMS level of the NVSS survey. 

Any source that is found in LoTSS but not in TGSS nor NVSS and has a flux density above 100 mJy, might still be an interesting transient candidate. Since we apply no compactness requirement in this search most of the candidates in our final sample show extended emission only. After visual inspection of the 25 LoTSS sources without TGSS nor NVSS counterpart we find one source that has a compact and extended component (ILT J183846.09+325110.7). We investigate this source by recalculating the contribution to integrated flux density of the point source and the extended emission separately and find that the compact part only has an integrated flux density of 18 mJy. The extended emission contributes to the total integrated flux density such that the combination crosses the 100 mJy limit. Due to the sensitivity difference of TGSS to LoTSS we do not necessarily expect to find this point source in TGSS. Furthermore, we reimage the LoTSS data with the TGSS baselines and weighting and find that the extended emission is not visible for TGSS. All in all, we conclude that the number of transient candidates missed due to inadequate source finding of a combination of point source and extended emission, is negligible.

\subsection{Variability \label{sec:variability}}
Although we focus on transient sources in this study, we can easily perform a simple variability search using Figure \ref{fig:scaled_flux_diff} by evaluating the tails of the distribution. Figure \ref{fig:scaled_flux_diff} shows the normalized flux density difference (as defined in Eqn. \ref{eqn:flux_diff}) for compact brighter than 100 mJy sources. We evaluate all matched sources where the flux density difference is larger than the median plus four times the standard deviation of the distribution, or smaller than the median minus four times the standard deviation. We find 7 sources that satisfy these criteria. These sources are listed in Table \ref{tab:variables}. Except for the first source, all of these variable candidates are sources where the TGSS flux density is below the completeness limit of 100 mJy. We therefore suspect that most of these sources suffer from similar effects as described in Section \ref{sec:results_LoTSS}, where poor TGSS image quality would significantly decrease the TGSS source flux density. Follow-up studies are needed to identify whether these sources are really variable. The only variable candidate where the flux significantly decreases from TGSS to LoTSS is source J014946+362832 (ID 1 in Table \ref{tab:variables}). The source type is unknown, but considering the flux density this variable source is most likely an AGN \citep{wilman2008semi, williams2016lofar, calistro2017lofar}.

\begin{figure*}
    \centering
    \includegraphics[width=\textwidth]{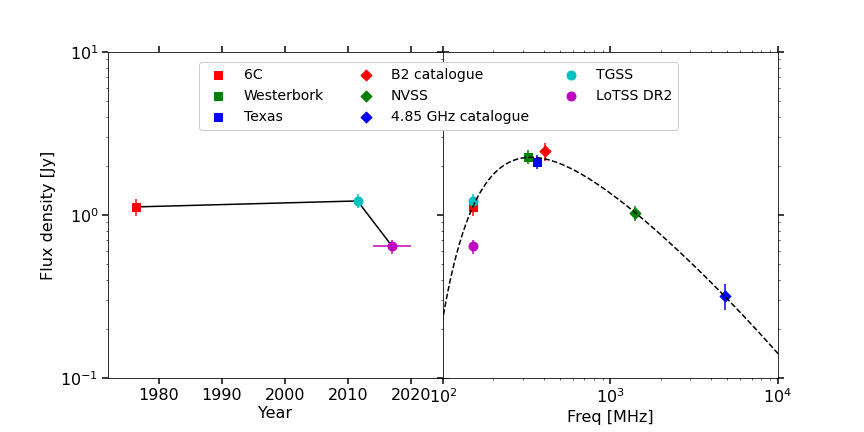}
    \caption{Light curves and spectrum of variable source J014946+362832 (ID 1 in Table \ref{tab:variables}). The light curve shows data at 151 MHz, including the 6C survey \protect \citep{hales19936c}, TGSS ADR1  \protect \citep{intema2017TGSSADR} and LoTSS DR2 ( \protect \citeauthor{shimwell2021DR2} in prep.). The spectrum furthermore includes data from the Westerbork survey  \protect \citep{rengelink1997westerbork}, the Texas survey  \protect \citep{douglas1996texas}, the B2 catalogue  \protect \citep{colla1973b2}, the NVSS catalogue  \protect \citep{condon1998nvss} and a 4.85 GHz survey  \protect \citep{becker1991new}. Following  \protect \cite{helmboldt2008radio} we fit the spectrum with $Y = A + BX + C \rm{exp}(DX)$ where $Y = \rm{log}(F)$ with F the flux in Jy and $X = \rm{log}(\nu)$ with $\nu$ the frequency in MHz. The fit parameters are A = 1.70,  B = -0.22,  C = -464.50 and D = -3.53. The TGSS and LoTSS data are not used for the fit.}
    \label{fig:var_lightcurves_spectrum}
\end{figure*}

\begin{figure}
    \centering
    \includegraphics[width=\linewidth]{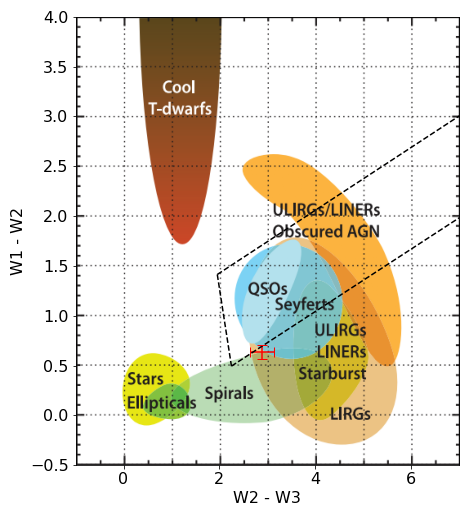}
    \caption{WISE infrared color-color space (W2 – W3 vs. W1 – W2) from \protect \cite{wright2010wide} where we overplot the variable source J014946+362832 (ID 1 in Table \ref{tab:variables}) in red using data from \protect \cite{cutri2014vizier}. The dashed black line shows the infrared color selection criteria for luminous AGNs in WISE defined by \protect \cite{mateos2012using}.}
    \label{fig:WISE_color}
\end{figure}

In Figure \ref{fig:var_lightcurves_spectrum} we show the light curve and the spectrum of this particular source. The light curve shows data at 151 MHz, including the 6C survey \citep{hales19936c}, TGSS ADR1 \citep{intema2017TGSSADR} and LoTSS DR2 (\citeauthor{shimwell2021DR2} in prep.). The spectrum also includes data from the Westerbork survey \citep{rengelink1997westerbork}, the Texas survey \citep{douglas1996texas}, the B2 catalogue \citep{colla1973b2}, the NVSS catalogue \citep{condon1998nvss} and a 4.85 GHz survey  \citep{becker1991new}. Following \cite{helmboldt2008radio} we fit the spectrum with $Y = A + BX + C \rm{exp}(DX)$ where $Y = \rm{log}(F)$ with F the flux in Jy and $X = \rm{log}(\nu)$ with $\nu$ the frequency in MHz. The fit parameters are A = 1.70,  B = -0.22,  C = -464.50 and D = -3.53. The TGSS and LoTSS data are not used for the fit. Follow-up studies are necessary to identify whether the overall intensity of the source has decreased or the spectrum has steepened towards low frequencies. The fact that the spectrum is peaked is indicative of an underlying absorption mechanism. These absorption mechanisms are often associated with source compactness due to youth of the source and/or confinement by an external medium \citep{o2021compact}. \cite{cutri2014vizier} provide WISE infrared data for source J014946+362832. In Fig \ref{fig:WISE_color} we show the location of the variable source in the WISE infrared color-color space (W2 – W3 vs. W1 – W2). The source location is plotted in red and interesting classes of objects are shown in the background (Figure 12 from \cite{wright2010wide}). Furthermore, the dashed black line shows the infrared color selection criteria for luminous AGNs in WISE defined by \cite{mateos2012using}. Assuming that the source is most likely an AGN, Fig. \ref{fig:WISE_color} shows that J014946+362832 could be a luminous infrared galaxy (LIRG), QSO or Seyfert. The figure also shows that the data point for J014946+362832 (including error bars) lies just at the edge of the region defining luminous AGN in WISE. More extensive (multi-wavelength) follow-up studies are required to identify the source type.

\subsection{Future strategies}
An improvement left for future work is a local noise dependent completeness limit. In this work we choose the completeness limit as described in \cite{intema2017TGSSADR} as a cutoff for the complete survey. However, there are differences in data quality and RMS noise throughout the survey. Therefore, a deeper transient search could be accomplished if one was to incorporate the local noise levels into this flux density cutoff. Low noise sky regions might yield a lower sensitivity than implied by the full-survey completeness limit and thus a deeper transient search. In contrast, for high noise regions a higher flux density cutoff might be applicable, reducing the number of false positives. \cite{carbone2016new} describe such a method in detail. To put such a strategy in place in future work it would be helpful if radio surveys provided a RMS map of the full survey area.

\section{Conclusions}
We present the results of a blind transient search at low frequency by comparing TGSS ADR1 and LoTSS DR2. We use the same universally applicable method to search for transients in TGSS that are not present in LoTSS and vice versa. The transient candidates we find are both imaging artefacts resulting from sidelobes of the point spread function at the location of bright sources and sources that are labelled as transient due to the large difference in flux density scales between the two surveys. We present methods to mitigate both these effects in future studies.

We conclude that there are no significant transient sources at a timescale of 2--9 years and a sensitivity of 100\,mJy (TGSS ADR1 completeness limit), which leads to an upper limit on the transient rate of $\rho < 5.4 \cdot 10^{-4}\ \text{deg}^{-2} $. This is an upper limit on the transient surface density at the longest timescale to date, using our pragmatic definition of a transient (Sect.~\ref{sec:methods_strategy}): a compact source that appears brighter than 100\,mJy in the catalogue of one survey, without a counterpart in the other survey. We note that a radio transient was detected by \cite{law2018discovery} at an even longer timescale (23 years) than the ones explored in this study, but is was detected a higher frequency of 1.4 GHz. Repeating this study with the final LoTSS data release, which will cover the entire Northern sky \citep{shimwell2017lofar}, will at least put an upper limit of $1.5 \cdot 10^{-4}\ \text{deg}^{-2}$ on the transient surface density at decade long timescale and a sensitivity of 100 mJy. However, an actual transient detection would provide more constraining values of the transient surface density. Finally, we estimate a maximum of one in $2.4 \cdot 10^3$ AGN to show extreme long-term variability at 150 MHz above a flux density of 100 mJy. In future studies a RMS map of the complete surveyed area would be a valuable asset to conduct possibly deeper transient studies (see \cite{carbone2016new}). 

\section*{Data availability statement} \label{sec:sup_material}
The data underlying this article are available in a reproduction package via Zenodo, at \url{https://dx.doi.org/10.5281/zenodo.4745528}

\section*{Acknowledgements}
IdR thanks Kelly Gourdji for fruitful discussions. GL thanks the Observatoire de la Côte d’Azur and the
Anton Pannekoek Institute for making this joint research work possible. AD acknowledges support by the BMBF Verbundforschung under the grant
05A20STA.

This research made use of the TOPCAT \citep{taylor2005topcat} pair match algorithm, Aladin \citep{bonnarel2000aladin} for visual verification, astropy \citep{astropy:2013, astropy:2018} for FITS file handling and matplotlib \citep{hunter2007matplotlib} was used to create plots. 

We thank the staff of the GMRT that made these observations possible. GMRT is run by the National Centre for Radio Astrophysics of the Tata Institute of Fundamental Research.

LOFAR is the Low Frequency Array designed and constructed by ASTRON. It has observing, data processing, and data storage facilities in several countries, which are owned by various parties (each with their own funding sources), and which are collectively operated by the ILT foundation under a joint scientific policy. The ILT resources have benefited from the following recent major funding sources: CNRS-INSU, Observatoire de Paris and Université d'Orléans, France; BMBF, MIWF-NRW, MPG, Germany; Science Foundation Ireland (SFI), Department of Business, Enterprise and Innovation (DBEI), Ireland; NWO, The Netherlands; The Science and Technology Facilities Council, UK; Ministry of Science and Higher Education, Poland; The Istituto Nazionale di Astrofisica (INAF), Italy.

This research made use of the Dutch national e-infrastructure with support of the SURF Cooperative (e-infra 180169) and the LOFAR e-infra group. The Jülich LOFAR Long Term Archive and the German LOFAR network are both coordinated and operated by the Jülich Supercomputing Centre (JSC), and computing resources on the supercomputer JUWELS at JSC were provided by the Gauss Centre for Supercomputing e.V. (grant CHTB00) through the John von Neumann Institute for Computing (NIC).

This research made use of the University of Hertfordshire high-performance computing facility and the LOFAR-UK computing facility located at the University of Hertfordshire and supported by STFC [ST/P000096/1], and of the Italian LOFAR IT computing infrastructure supported and operated by INAF, and by the Physics Department of Turin university (under an agreement with Consorzio Interuniversitario per la Fisica Spaziale) at the C3S Supercomputing Centre, Italy.

This research was funded through project CORTEX (NWA.1160.18.316), in research programme NWA-ORC, financed by the Dutch Research Council (NWO).




\bibliographystyle{mnras}
\bibliography{LoTSSvTGSS} 


\appendix
\section{TGSS imaging artefacts} \label{app:TGSS_imaging_artefacts}
The upper part of Table \ref{tab:artefact_list} shows the ten candidates that are left after visual inspection following the steps in Section \ref{sec:methods_strategy} as mentioned in the second row of Table \ref{tab:nof_sources_per_step}. These candidates have an integrated flux density above 100 mJy and are compact. Other sources in Table \ref{tab:artefact_list} were found in a follow-up, lowering the flux density threshold to 70 mJy and dropping the compactness requirement. TGSS and LoTSS images of all brighter than 100 mJy candidates can be found in the reproduction package.

\begin{table*}
\caption{TGSS to LoTSS transient candidates identified as imaging artefacts. This table shows for each transient candidate the transient candidate source name, the transient candidate source flux density, the source name of the bright close by source, the source flux density of this bright source, the distance between the transient candidate and the bright source and whether or not the transient candidates meets our compactness criteria (as discussed in section \ref{sec:methods_compactness} and Figure \ref{fig:compactness_combined}). The ten candidates with an integrated transient flux density brighter than 100 mJy (upper half of the table) which are compact as well are the 10 transient candidates as initially mentioned in Section \ref{sec:methods_strategy} and Table \ref{tab:nof_sources_per_step}. The other candidates in the table below are found after a follow-up where we decrease the flux density threshold to 70 mJy and drop the compactness criterion. \label{tab:artefact_list}}
\begin{tabular}{ccccccc}
  \hline
 & ‘Transient’ candidate source name & Flux density (mJy) & Close by bright source name & Flux density (Jy) & Distance ($\arcmin$) & Compact \\ \hline
1 & TGSSADR J122030.3+334531 & 838.0 & TGSSADR J122033.7+334311 & 20.1 & 2.45 & no \\
2 & TGSSADR J123649.2+365757 & 350.6 & TGSSADR J123649.8+365517 & 6.8 & 2.66 & yes \\
3 & TGSSADR J151330.6+472410 & 286.1 & TGSSADR J151322.2+472150 & 4.6 & 2.74 & no \\
4 & TGSSADR J130028.6+400541 & 237.1 & TGSSADR J130033.0+400907 & 12.3 & 3.55 & yes \\
5 & TGSSADR J124447.9+361156 & 230.6 & TGSSADR J124449.5+360924 & 5.1 & 2.55 & no \\
6 & TGSSADR J154824.1+483721 & 220.6 & TGSSADR J154814.5+483501 & 9.3 & 2.83 & no \\
7 & TGSSADR J160454.1+573014 & 214.8 & TGSSADR J160434.3+572801 & 4.9 & 3.46 & yes \\
8 & TGSSADR J215844.6+295639 & 207.5 & TGSSADR J215842.0+295908 & 6.7 & 2.54 & no \\
9 & TGSSADR J145418.4+500554 & 196.6 & TGSSADR J145408.2+500331 & 4.0 & 2.89 & no \\
10 & TGSSADR J155957.2+533830 & 182.8 & TGSSADR J160016.7+533944 & 6.3 & 3.14 & yes \\
11 & TGSSADR J094940.0+661228 & 175.5 & TGSSADR J094912.2+661500 & 4.3 & 3.77 & yes \\
12 & TGSSADR J094218.2+602247 & 155.3 & TGSSADR J094151.2+602048 & 5.4 & 3.88 & no \\
13 & TGSSADR J003233.5+195607 & 152.0 & TGSSADR J003238.3+195353 & 4.1 & 2.50 & no \\
14 & TGSSADR J152514.9+533637 & 146.9 & TGSSADR J152501.8+533411 & 2.8 & 3.11 & yes \\
15 & TGSSADR J150929.5+472914 & 141.6 & TGSSADR J150920.0+472655 & 2.6 & 2.82 & no \\
16 & TGSSADR J125720.6+364641 & 141.0 & TGSSADR J125723.7+364418 & 4.0 & 2.47 & yes \\
17 & TGSSADR J232131.5+295504 & 132.2 & TGSSADR J232143.9+295542 & 4.4 & 2.76 & no \\
18 & TGSSADR J140012.5+533936 & 129.1 & TGSSADR J140018.9+533659 & 6.0 & 2.78 & yes \\
19 & TGSSADR J145812.7+483518 & 118.5 & TGSSADR J145802.0+483304 & 2.3 & 2.85 & yes \\
20 & TGSSADR J154533.0+462506 & 108.2 & TGSSADR J154525.3+462244 & 5.2 & 2.70 & yes \\ \hline
21 & TGSSADR J115905.7+535541 & 97.3 & TGSSADR J115913.7+535307 & 7.1 & 2.82 & yes \\
22 & TGSSADR J161156.3+552133 & 95.6 & TGSSADR J161212.5+552305 & 3.9 & 2.75 & yes \\
23 & TGSSADR J145601.0+474314 & 92.8 & TGSSADR J145551.0+474056 & 3.1 & 2.85 & yes \\
24 & TGSSADR J010257.2+255016 & 92.6 & TGSSADR J010250.1+255216 & 5.3 & 2.56 & yes \\
25 & TGSSADR J012832.5+290615 & 89.0 & TGSSADR J012830.1+290300 & 13.7 & 3.30 & yes \\
26 & TGSSADR J011128.0+260317 & 86.9 & TGSSADR J011121.3+260518 & 3.4 & 2.51 & yes \\
27 & TGSSADR J153124.0+353739 & 86.4 & TGSSADR J153125.2+353340 & 14.1 & 3.99 & yes \\
28 & TGSSADR J004105.5+330653 & 83.5 & TGSSADR J004054.9+331006 & 14.3 & 3.91 & yes \\
29 & TGSSADR J124011.0+350516 & 82.5 & TGSSADR J124021.1+350258 & 1.8 & 3.10 & yes \\
30 & TGSSADR J133003.0+453824 & 81.6 & TGSSADR J132942.1+453957 & 2.2 & 3.96 & yes \\
31 & TGSSADR J132343.5+411344 & 81.5 & TGSSADR J132323.9+411514 & 3.9 & 3.98 & yes \\
32 & TGSSADR J152758.3+514444 & 78.1 & TGSSADR J152746.7+514226 & 3.0 & 2.92 & yes \\
33 & TGSSADR J225715.9+170059 & 77.2 & TGSSADR J225707.1+165822 & 10.5 & 3.35 & yes \\
34 & TGSSADR J003601.5+184017 & 77.1 & TGSSADR J003606.4+183759 & 13.4 & 2.58 & yes \\
35 & TGSSADR J160357.5+572931 & 76.2 & TGSSADR J160434.3+572801 & 4.9 & 5.18 & yes \\
36 & TGSSADR J100546.4+345530 & 73.9 & TGSSADR J100601.7+345409 & 9.6 & 3.40 & yes \\
37 & TGSSADR J004636.1+285054 & 72.6 & TGSSADR J004642.9+284850 & 2.4 & 2.56 & yes \\
38 & TGSSADR J154156.2+525316 & 71.8 & TGSSADR J154144.7+525054 & 2.1 & 2.92 & yes \\
39 & TGSSADR J093418.6+671828 & 70.7 & TGSSADR J093346.4+672052 & 2.9 & 3.93 & yes \\
40 & TGSSADR J084001.0+400146 & 70.7 & TGSSADR J084011.4+400348 & 4.5 & 2.84 & yes \\
41 & TGSSADR J154546.1+482919 & 70.0 & TGSSADR J154535.8+482703 & 1.4 & 2.84 & yes \\
 \hline 
 \end{tabular}
\end{table*}

\section{Variable sources}
Table \ref{tab:variables} shows a list of the variable sources identified via the methods described in Section \ref{sec:variability}.

\begin{table*}
\caption{Compact TGSS sources that are matched to a compact LoTSS source where the normalized flux density difference (Eqn. \ref{eqn:flux_diff}) is larger than the median plus 4 times the standard deviation or smaller than the median minus 4 times the standard deviation of the distribution shown in \ref{fig:scaled_flux_diff}. These are outliers are variable sources. The spectrum and light curve of source ID 1 are shown in Fig. \ref{fig:var_lightcurves_spectrum}. The columns show an ID,  the normalized difference between the LoTSS and TGSS flux density difference, the TGSS source name and flux density, the LoTSS source name and flux density and the RA and DEC coordinates. 
\label{tab:variables}}
\begin{tabular}{cc cc cc cc}
  \hline
ID & Normalized diff & TGSS source name & Flux (mJy) & LoTSS source name & Flux (mJy) & RA ($\deg$) & DEC ($\deg$) \\ \hline
1 & -4.19 & TGSSADR J014946.0+362832 & 1218.1 & ILT J014946.05+362832.6 & 641.0 & 27.442 & 36.476 \\
2 & 6.31 & TGSSADR J023024.1+283520 & 28.7 & ILT J023024.02+283521.5 & 117.1 & 37.600 & 28.589 \\
3 & 5.59 & TGSSADR J220546.1+292655 & 52.5 & ILT J220546.51+292655.1 & 154.2 & 331.444 & 29.449 \\
4 & 5.49 & TGSSADR J144452.7+630712 & 55.1 & ILT J144452.35+630711.6 & 146.9 & 221.218 & 63.120 \\
5 & 5.15 & TGSSADR J181549.8+382450 & 61.2 & ILT J181549.94+382450.2 & 153.7 & 273.958 & 38.414 \\
6 & 5.84 & TGSSADR J132707.5+342336 & 39.1 & ILT J132707.56+342337.9 & 114.3 & 201.782 & 34.394 \\
7 & 5.05 & TGSSADR J175246.4+355037 & 49.9 & ILT J175246.50+355036.6 & 131.2 & 268.194 & 35.844 \\
 \hline 
 \end{tabular}
\end{table*}



\bsp	
\label{lastpage}
\end{document}